\documentclass[%
onecolumn,%
oneside,%
floats,%
aps,%
 prd,%
nobibnotes,%
nofootinbib,%
amsmath,%
amssymb,%
amsfonts,%
amscd,%
   showpacs,%
  superscriptaddress,%
 flushb]
 {revtex4}

\usepackage[utf8]{inputenc}
\usepackage{graphicx,array,dcolumn}
\usepackage{cases}
\usepackage[paperwidth=210mm,paperheight=297mm,centering,hmargin=1.8cm,vmargin=2.5cm]{geometry}
\usepackage{enumerate}

\DeclareGraphicsExtensions{.pdf,.png,.jpg}
\usepackage{subcaption}

\usepackage{color}

\usepackage{hyperref}

\newcommand{\bi}{\bibitem}

\newcommand{\rar}{\rightarrow}

\newcommand{\be}{\begin{equation}}
\newcommand{\ee}{\end{equation}}

\def\be{\begin{eqnarray}}
\def\ee{\end{eqnarray}}

\def\-g{\sqrt{-g}}

\renewcommand\rho{\varrho}

\begin{document}


\title{Cosmic rays from heavy particle decays}

\author{E.V. Arbuzova}
\email{arbuzova@uni-dubna.ru}
\affiliation{Novosibirsk State University, Novosibirsk, 630090, Russia}
\affiliation{Department of Higher Mathematics, Dubna State University, 141982 Dubna, Russia}

\author{A.D. Dolgov}
\email{dolgov@nsu.ru}
\affiliation{Novosibirsk State University, Novosibirsk, 630090, Russia}
\affiliation{Bogoliubov Laboratory of Theoretical Physics, JINR, Dubna, 141980, Russia}

\author{A.A. Nikitenko}
\email{nikitenko@theor.jinr.ru}
\affiliation{Bogoliubov Laboratory of Theoretical Physics, JINR, Dubna, 141980, Russia}
\affiliation{Department of Fundamental Problems of Microworld Physics, Dubna State University, Dubna, 141982, Russia}

\begin{abstract} 
Multidimensional modification of gravity with a smaller mass scale of the gravitational interaction is
considered. Stable by assumption dark matter  particles could decay {via interactions} with 
virtual black holes. The decay rates of such processes are estimated.
It is shown that with the proper fixation of the parameters the decays of these ultra-massive particles   
can give noticeable contribution to the flux of high energy cosmic rays 
in particular,  near the Greisen-Zatsepin-Kuzmin limit. Such particles can also create
neutrinos of very high energies observed in the existing huge underwater or ice-cube detectors.
\end{abstract}

\maketitle

\section{Introduction}

The problem of the origin of the ultra high energy cosmic rays (UHECR) is one of the most important problems in cosmic 
ray physics. Especially interesting are fluxes of particles with energies close to the 
Greisen-Zatsepin-Kuzmin (GZK) limit~\cite{Greisen:1966jv,Zatsepin:1966jv}, $E \gtrsim 10^{20}$~eV,
 sometimes called Extremely High Energy Cosmic Rays, abbreviated as EHECR.
According to the quoted papers the mean free pass of protons due to scattering on the 
cosmic microwave background radiation is about 50 Mpc. So one is forced to conclude that the sources of UHECR should be sufficiently close to the solar system. 
However, no proper astrophysical sources are observed in our neighbourhood.   

There is no single opinion on the origin of the ultra high energy cosmic rays. The traditional approach is that the high energy cosmic rays are 
created by astrophysical sources such as active galactic nuclei, Seyfert galaxies, or possibly hypernovae. For a recent review see Ref. \cite{Kachelriess:2022phl}.
Another interesting possibility is the cosmic ray emission by topological defects, reviewed in Ref.~\cite{Sigl:1996ak}. More recent discussion can be found in 
Ref.~\cite{Semikoz:2007wj}.  Superheavy particle decays are also considered as a possible source of energetic cosmic rays 
\cite{Berezinsky:1997hy,Kuzmin:1997jua,Birkel:1998nx}. It is assumed that there exist superheavy dark matter (DM) particles with masses in the range
$(10^{13} -10^{16})$~GeV and with the life-time significantly larger than the universe age. 
The required instability is achieved either by a small breaking of 
$R$-parity in supersymmetric theories, or instanton and wormholes effects. 
{An essential clarification of the mechanism of UHECR production through heavy 
particle decays and the corresponding phenomenology was achieved in papers \cite{kt1,kt2}.}

{There are two distinct ranges in energy of UHECR. Cosmic rays with $E \lesssim 10^{20} $ eV  
may be created by stellar processes by acceleration of stellar material in catastrophic processes. The cosmic rays 
with such energies naturally contain a large amount of nuclei. Cosmic nuclei with higher energies, 
$E \gtrsim 10^{20} $ eV  is difficult to create in stellar processes. In principle, such ultra energetic cosmic rays
could be produced in heavy particle decays. In this case they would not contain nuclei but only stable elementary
particles, i.e. only protons and photons. Even the most long-lived elementary particle, muon, cannot reach
the Earth in significant numbers. }

In the present work we study a possibility that the source of UHECR {at energies $E \gtrsim 10^{20} $ eV could  be} decays of superheavy dark matter particles, {while the cosmic rays with lower energies could be
created through astrophysical processes.}
{ The adopted here mechanism of instability is
completely different from those considered in the literature}.

In what follows we denote these {superheavy DM}  particles as $X$-particles. 
{ For the mechanism of $X$-particle production we do not suggest anything new but simply use that
based on the $R^2$ inflation~\cite{Starobinsky:1980te},
 studied in our papers~\cite{Arbuzova:2021etq,Arbuzova:2021oqa}, as well as in several other ones.
 This mechanism possesses  some specific features facilitating heavy particle
 production.}
However, with the canonical energy scale of gravitational interaction equal to 
$M_{Pl} = 1.22 \times 10^{19}$ GeV, the life-time of $X$-particles turns out to be too long to allow for
any observable  consequences of their decays.  A possible way out could be opened by diminishing the 
fundamental gravity scale at small distances down to a lower value $M_* < M_{Pl}$. 
This could lead to a considerable increase of decay probabilities of $X$-particles. 

{To describe mutual interactions in the world of superheavy particles we assume, though it is not obligatory,
an existence of supermassive quarks that interact through a new QCD-like force. 
This force binds heavy quarks analogously to the
usual low energy QCD but at much higher energy scale. Our task is to fix the values of parameters and in particular
the value of the new QCD scale $\Lambda_*$ to allow for a proper value of the life-time of dark matter particles $X$. 
As one can see below, we need to this end $\Lambda_* \sim 10^{15}$ GeV.
}

{We are interested in the value of $\Lambda_*$ when the QCD
coupling constant $\alpha_s$, becomes equal unity. 
According to Figs. 3.1 and 3.2 of the review~\cite{QCD-rev} the QCD coupling constant
$\alpha_s$ reaches unity at  the momentum transfer $Q^2$ close to $\Lambda^2$ practically at the 
same values of $\Lambda$. This is enough for our order of magnitude estimates. 
This shows that the mechanism suggested in our paper in principle may operate. Fig. 3.4 
demonstrates that the results for $\alpha_{s}$ weakly depends upon the renormalization schemes. }

Reducing the fundamental gravity scale can be naturally 
achieved assuming multidimensional modification of gravity, as it is studied in Ref.~\cite{Bambi:2006mi}.
The results of this work are not directly applicable to our purpose since  we are interested in the mass
scale of the DM particles about $10^{12}$ GeV, so that their decays could contribute to the flux of UHECR.
On the other hand,  in paper~\cite{Bambi:2006mi} unification of gravity with electroweak 
interactions at TeV-scale, as suggested in Refs.~\cite{Arkani-Hamed:1998jmv,Antoniadis:1998ig}, was studied and the results obtained in~\cite{Bambi:2006mi} 
could be extended up to the scale of the order about $10^{12}$ GeV.   
Multidimensional modification of gravity in somewhat different frameworks was considered also in several papers, see e.g. 
\cite{Nikulin:2022ryt,Petriakova:2022esq} and references therein.

Usually dark matter particles are supposed to be {absolutely} stable. However, there exists a mechanism suggested by Ya.~B.~Zeldovich ~\cite{Zeldovich:1976vq,Zeldovich:1977be}, which 
leads to decay of any presumably stable particles through creation and evaporation of virtual black holes.  
However, the rate of the proton decay calculated in the canonical gravity, 
{with the energy scale equal to $M_{Pl}$,} is extremely tiny. 
It is shown in what follows that smaller scale of gravity and huge mass of DM particles {both}
lead to a strong amplification of the Zeldovich effect.

{ Heavy $X$-particles decay into all kinds of elementary particles. However, only stable ones, such as protons and photons, 
could reach detectors on the Earth. In our model mesons and leptons are produced at the point of the
heavy particle decay, but only stable ones created by those mesons and leptons can be registered by the detector.  

The most long-lived particle, muon, may reach the detector if the decay of superheavy particles took place 
very close to the Earth and so the contribution of muons to the flux of cosmic rays is negligible, because of small 
total number of superheavy particles at the muon decay distance, even if gamma-factor of the muon decay is taken into 
account.

An interesting signature of $X$-particle contribution into EHECR could be an anisotropy of the flux. 
In our scenario the dominant contribution  comes to the highest energy cosmic rays only. For these energies statistics is too small to make 
definitive conclusion on presence or absence of anisotropy. Still some indications to an anisotropy are observed, see e.g.
Ref. \cite{TelescopeArray:2021dpk}. 
} 

The paper is organised as follows. In Section \ref{s-decay-BH} the basics of the Zeldovich mechanism, 
the conditions of absence of naked singularity created by the virtual black hole, and the parameters of multidimensional gravity are discussed. 
In Section \ref{s-p-type-DM} the  decay rate of the heavy proton-type dark matter particle through virtual BH 
is calculated and the value of the parameters are fixed in such a way that the life-time of this particle is $3-4$ orders of magnitude longer 
than the universe age.
In Section \ref{s-l-type-DM} we assume that the DM particle is a superheavy lepton and evaluate the probability of its decays.   
In Section \ref{s-R2-modified} we argue that superheavy leptons could be produced in $R^2$-modified gravity and fix the values of the scalaron mass 
and of the gravity scale in $(d+4)$-dimensional space.

\section{Decays through virtual black holes \label{s-decay-BH} }
 
It is known that black holes break all global symmetries. In particular, the violation of 
the global $U_B(1)$ and $U_L(1)$ would lead to baryonic and leptonic numbers non-conservation. As it 
{has been argued} 
in Refs.~\cite{Zeldovich:1976vq,Zeldovich:1977be},  
proton would become unstable due to interaction {of its constituents, three light quarks,
with a virtual black hole.} 
Following Ref.~\cite{Bambi:2006mi} we consider the process of transformation of  two quarks  inside proton into antiquark and antilepton through a virtual black hole:
\be
q + q \rightarrow (BH) \rightarrow  \bar{q} + \bar{l}.
\label{qq-barql}
\ee  
The emerging antiquark annihilates with the remaining third quark in proton, {e.g. into pions,}
inducing proton decay. This {last} process is practically instantaneous and thus it does not have 
noticeable impact on the proton life-time.  

In accordance with Zeldovich works \cite{Zeldovich:1976vq,Zeldovich:1977be}, we evaluate the probability of this process  per unit time as:
\be
\dot n /n = n \sigma_{BH}  = \sigma_{BH} |\psi (0)|^{2} ,
\label{dotn-n}
\ee
where $n\sim m_p^3$ is the number density of quarks inside the proton, $m_p$ is the proton mass, 
$\sigma_{BH}$ is the cross-section of process  \eqref{qq-barql}, and $ \psi $ is the wave function of three quarks inside proton.

{The presented above estimate of the supermassive quarks density inside proton, $n\sim m_p^3$,
deserves some explanation. For example the size of the bound state of two weakly interacting particles, e.g. 
positronium (or muonium) is of the order $1/(\alpha m$), to be more precise $2/(\alpha m$), where $m$ is the electron mass. 
Here $\alpha = 1/137 \ll 1$. For strongly bound systems with the effective coupling constant $\alpha_s$ equal unity the size is  of the 
order of $1/m$. However, there is an interesting exception, namely pion that is anomalously light. The reason is that pion is a 
pseudogoldstone boson and in the case of unbroken chiral symmetry the pion would be massless.
}

Using dimensional consideration we approximate this cross-section as:
\be
\sigma_{BH} \sim m_p^2/M_{Pl}^4 .
\label{sigma-BH}
\ee
Expressions \eqref{dotn-n} and \eqref{sigma-BH} lead to the following result for the proton life-time:
\be \label{tau-p}
\tau_p= \frac{n}{\dot n} = \frac{M_{Pl}^4}{m_p^5}. 
\ee 
Taking the value of the Planck mass, $M_{Pl} = 1.22 \times10^{19}$GeV $= 2.18 \times 10^{-5}$g, we restore the result of Refs.~\cite{Zeldovich:1976vq,Zeldovich:1977be}, 
$\tau_p \sim 10^{45}$ years. This is well inside the existing experimental lower bound on the proton life-time $10^{31}$ to $10^{33}$ years depending upon the decay 
mode \cite{ParticleDataGroup:2022pth}.

If we make  {the estimate} 
analogous to \eqref{tau-p} for heavy $X$-particles with mass $M_X = 10^{6}$ GeV 
we {find} that their life-time is $\tau_X^{(6)} \sim 10^{14}$ years, which is 
safely larger that the universe age, $t_U \approx 1.5 \times 10^{10}$ years 
{and allows for a noticeable conribution to the flux of cosmic rays at the energies $10^6$ GeV.
However, if we need to explain the violation of the GZK bound we need $M_X \sim 10^{12}$ GeV.
In this case the life-time of  $X$-particle would be much shorter, }
$\tau_X^{(12)} \sim 1.5\times10^{-8}$ sec.   
{Such a short life-time  does not permit these 
particles to act as dark matter.}

The life-time would be significantly {{increased}, 
if virtual black holes satisfy the same restrictions that valid for classical black holes. 
{Namely, these black holes should be devoid of any conserved quantum numbers,}
i.e. they should have zero electric and 
colour charges (as well as any other local charges), as argued in Ref.~\cite{Bambi:2006mi}. Indeed, the horizon radius, $R_{BH}$, of the Kerr--Newman black hole 
with mass $M_{BH}$, electric charge $Q$ and angular momentum $J$ is  equal to \cite{Misner:1973prb,lppt}:
\be
R_{BH} M_{Pl} = \frac{M_{BH}}{M_{Pl}} + 
\sqrt{\left( \frac{M_{BH}}{M_{Pl}}\right)^2 - Q^2 - J^2  
\left(\frac{M_{BH}}{M_{Pl}}\right)^{-2}}.
\label{r-hor}
\ee 
If $M_{BH}<M_{Pl}$, as it is true in the case under consideration, there generally appears the forbidden naked singularity if $Q$ and $J$ are nonzero. 
Consequently the formation of such black holes is {not allowed} 
and we should look for more complicated processes leading to creation of virtual black holes 
with vanishing electric charge and angular momentum, {as is suggested in Ref.~\cite{Bambi:2006mi}. }
For example the proton decay can be described by the diagram presented in Fig. \ref{f:h-dt}. 

\vspace{1.5cm}
	\begin{figure}[h]
	\vspace{-5cm}
  	\centering
  		\begin{minipage}[b]{0.9\textwidth}
  		\includegraphics[width=\textwidth]{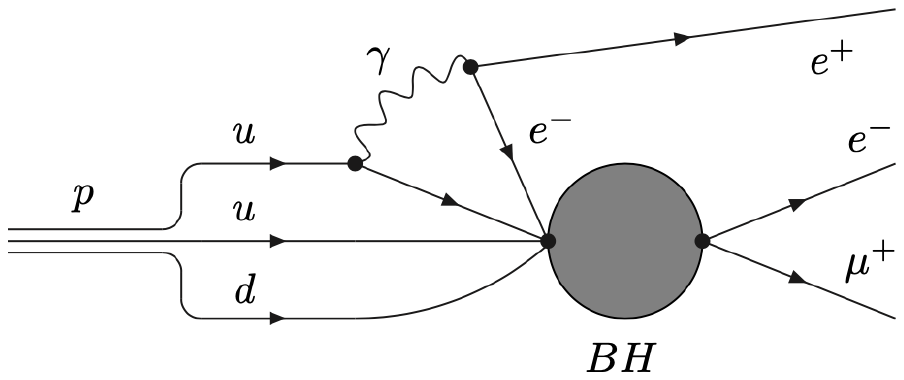}    
		 \end{minipage}
		 \vspace{-3cm}
		 \caption{{Diagram describing gravitationally induced proton decay, according to Ref.~\cite{Bambi:2006mi}.}}
		 \label{f:h-dt}
	\end{figure}
An essential feature of this graph is that neutral and 
spinless black hole may be formed only from a multi-particle initial state. This condition leads to a strong 
suppression of the decay probability. 

 {According to calculations of Ref. \cite{Bambi:2006mi}  the width of the proton decay
 into positively charged lepton and quark-antiquark pair is:} 
\be 
\Gamma(p\rar l^+ \bar q q) = 
\frac{m_p\,\alpha^2}{ 2^{12} \, \pi^{13}}
\left(\ln \frac{M_{Pl}^2}{m_q^2}\right)^2 \,
\left(\frac{\Lambda}{M_{Pl}}\right)^6 \,
\left(\frac{m_p}{M_{Pl}}\right)^{4+\frac{10}{d+1}}\, 
\int_0^{1/2} dx x^2 (1-2x)^{1+\frac{5}{d+1}},
\label{gamma-p}
\ee
 where $m_p \approx 1$GeV is the proton mass, $m_q \sim 300$ MeV is the constituent quark mass,  
 $\Lambda \sim 300$ MeV is the QCD scale parameter, $\alpha = 1/137$ is the fine structure constant,
 and $d$ is the number of "small' extra dimensions. {The QCD coupling constant $\alpha_s$ is supposed to be equal to unity.}
 {We can check that the proton decay rate is extremely small 
and the corresponding life-time is $7.3\times 10^{198} $ years which is
by far longer than the universe age, $t_U \approx 1.5\times 10^{10}$  years.}
 
 This case of decaying proton is mentioned for illustration only. We are interested in superheavy DM
 particles with masses about $10^{12}$ GeV and trying to formulate the scenario leading to their
 life-time with respect to the decay through the virtual BH only a few orders of magnitude longer than
 the universe age. Decays of such particles could make essential contribution to the UHECR spectrum.
 We argue that it can be achieved in the theory, where  gravitational coupling rises up at small distances 
 or high energies.

An example of such theory was proposed in Refs. \cite{Arkani-Hamed:1998jmv,Antoniadis:1998ig}. According to these works
 the observable universe with the Standard Model particles  is confined to 
  a 4-dimensional brane embedded in a
(4+$d$)-dimensional bulk, while gravity  propagates throughout the
bulk.  In such scenarios, the Planck mass $M_{Pl}$ becomes an
effective long-distance 4-dimensional parameter and the relation with
the fundamental gravity scale $M_{\ast}$ is given by
\begin{eqnarray}
M_{Pl}^2\sim M_{\ast}^{2+d}R_*^d ,
\label{M-ast}
\end{eqnarray}
where $R_*$ is the size of the extra dimensions: 
\be
R_{*} \sim \frac{1}{M_*}\left(\frac{M_{Pl}}{M_*}\right)^{2/d}.
\label{R-of-M}
\ee
For future application we choose 
{$M_* \approx 3\times10^{17}$ GeV, see below Eq.~(\ref{M-star})}
so $R_*\sim 10^{(8/d)}/M_* > 1/M_*$.

 The Schwarzschild radius of the virtual black hole created in the course of the proton decay is about $m_p/M_*^2 \ll R_*$, correspondingly the effective gravitational constant  is determined by $M_*$. Indeed, at the distances larger than $R_*$ the strength of gravity is determined by the canonical 
 value of gravitational coupling, $G_N = 1/M_{Pl}^2$, while at the distances smaller than $R_*$ the gravitational coupling is {given} by $M_*$:  $G_N^* = 1/M_*^2$.
 
 {\section{Heavy proton type dark matter \label{s-p-type-DM}}}
  
 Using Eq. \eqref{gamma-p} {with the substitution $M_*$ instead of $M_{Pl}$} and keeping the same 
 values of other parameters, we can estimate the proton life-time with respect to decay 
 $p\rar \bar q q l^+ $ 
 { for $M_* \approx 3\times10^{17}$ GeV, $\tau = 2.17 \times 10^{188}$ years.} 
 
 {However, we take as an example of DM particle
 a much heavier analogue of proton, namely the particles with the mass $10^{12}$ GeV, 
 that is stabilised, up to the virtual BH effects, by the approximate conservation of the 
 baryonic number, or something similar to it.
 The life-time of this particle is presented  below in eq. (\ref{tau-X}) as a function of the unknown
parameter $\Lambda_*$.} 

{We assume that these}
heavy dark matter particles with mass $M_X \sim10^{12}$ GeV consists of three heavy quarks, $q_*$, with comparable mass, 
{leaving $\Lambda_* $ as a free parameter.}
The life-time of $X$-particles can be evaluated using Eq.  \eqref{gamma-p} where we substitute 
$\alpha_*= 1/50$ instead of $\alpha = 1/137$, $M_X=10^{12}$ GeV  
instead of $m_p$, the mass of the constituent quark $m_{q_*} = 10^{12}$~GeV,
  and $d=7$: 
\be \label{tau-X}
\tau_X = \frac{1}{\Gamma_X} 
\approx {6.6\times 10^{-25} \rm{s} \, \cdot\frac{2^{10} \pi^{13}}{\alpha_*^2}} \left(\frac{\rm{GeV}}{M_X}\right)
\left(\frac{M_*}{\Lambda_*}\right)^6  \left(\frac{M_*}{M_X}\right)^{4+\frac{10}{d+1}}
\left( \ln \frac{M_*}{m_{q_*}} \right)^{-2} I(d)^{-1},
\ee
where we took {1/GeV = $6.6\times 10^{-25} $\rm{s}} and

\be
I (d) =\int_0^{1/2} dx x^2 (1-2x)^{1+\frac{5}{d+1}}, \,\,\, I(7) \approx 0.0057.
\label{i-of-d}
\ee
Now all the parameters, except for $\Lambda_*$, are fixed: $M_*=3\times 10^{17}$~GeV,
$M_X = 10^{12}$ GeV, $m_{q_*} \sim M_X$ and the life-time of X-particles can be estimated as: 
\be {
\tau_X \approx 7\times 10^{12}\,\,  {\rm years} \left(10^{15}\,\rm{GeV} /\Lambda_* \right)^6 .}
\label{tau-x-2}
\ee
A slight variation of $\Lambda$ near {$10^{15}$ GeV} allows to fix the life-time of the dark matter 
X-particles in the interesting range. They would be stable enough to behave as the cosmological dark 
matter and their decay could make considerable contribution into cosmic rays at ultra high energies. 

\vspace{1cm}

 {\section{Heavy lepton type dark matter \label{s-l-type-DM}}}

More promising and less subject to theoretical uncertainties is the hypothesis of
making dark matter from superheavy neutral leptons, $L$. 
Such leptons can be naturally stable with respect to particle physics interactions
due to leptonic number conservation if the heavy lepton, $L$, is not
mixed with lighter lepton species of the standard particle physics model.
This is analogous to the discussed in the literature stability of the 4th generation 
heavy neutrinos if the mixing with lighter first three generations vanishes.  

However, this assumption leads to 
 absolute stability of sterile lepton, because it would not decay
via  virtual black hole since $L$ has spin 1/2 
and cannot turn into
spinless particle which may be captured by the BH. To this 
end it is necessary to introduce some new interaction of $L$ with other elementary particles not destroying
its (quasi)stability.

The process of possible $L$ decay through a virtual  BH is described by the diagrams  similar to those 
describing muon decays presented in Figs.~~\ref{mu-3e} and \ref{mu-e-gamma}.
\begin{figure}[h]
	\centering 
	\vspace{-7cm} 
	\begin{minipage}[h]{0.8\textwidth}
		\includegraphics[width=\textwidth]{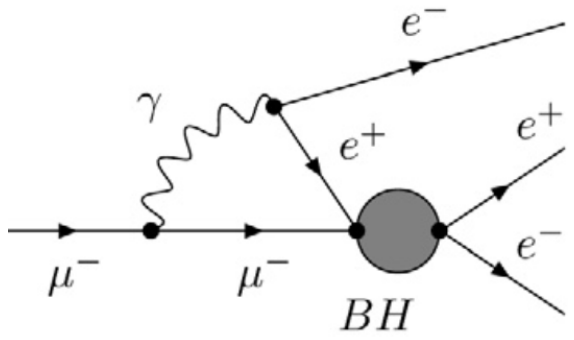}    
	\end{minipage}
	\vspace{-7cm}
	\caption{{Diagram describing $\mu \rar 3e$ 
			decay, according to Ref.~\cite{Bambi:2006mi}.}}
	\label{mu-3e}
\end{figure}

 {According to the calculations of  Ref.~\cite{Bambi:2006mi} the width of $\mu \rar 3e$ 
decay is:
\be
\Gamma(\mu\rar 3e)_d = \frac{\alpha^2 m_\mu}{2^{11}\pi^5}\,
\left(\ln \frac{M_*^2}{m_\mu^2}\right)^2\,
\left(\frac{m_\mu}{M_*}\right)^{4(1+\frac{1}{d+1})}
\kappa^{\frac{2}{d+1}}, 
\label{gamma-mu-3e}
\ee
where $m_\mu $ is the muon mass and coefficient $\kappa $ is approximately 1/2.
In the case of the ordinary (3+1)-dimensional gravity with $M_* =M_{Pl}$,
the decay rate is roughly equal to
\be
\Gamma (\mu\rar 3e)_{4D} \sim
{0.0015}\left(\frac{\alpha}{2\pi}\right)^2\, \frac{m_\mu^9}{M_{Pl}^8}
\label{mu-3e-4dim}
\ee 
and is extremely small. }

{As it is argued in Ref.~\cite{Bambi:2006mi}
the decay rate $\mu\rightarrow e\gamma$ can be estimated from diagram~\ref{mu-e-gamma}.
The probability  of the decay is suppressed by an additional power of $\alpha$, but the ratio of
two-body to three-body phase space compensates this extra suppression,
so that the probability of $\mu\rightarrow e\gamma$ decay through the
considered mechanism would be approximately the same as $\mu\rightarrow 3e$. }

 \begin{figure}[h]
 	\centering
 	\vspace{-7.5cm}
 	\begin{minipage}[h]{0.8\textwidth}
 		\includegraphics[width=\textwidth]{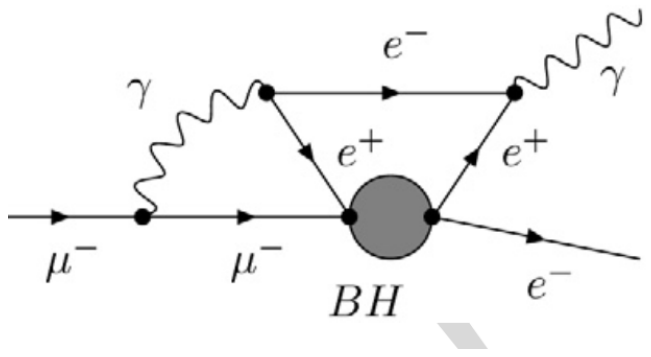}    
 	\end{minipage}
 	\vspace{-7cm}
 	\caption{{Diagram describing $\mu \rar e\gamma$ 
 			decay, according to Ref.~\cite{Bambi:2006mi}.}}
 	\label{mu-e-gamma}
 \end{figure} 
 
However, as we have mentioned above, these results are not directly applicable to  
the considered  case since we look for almost
sterile dark matter particle (lepton $L$), stable with respect to canonical particle channels. If 
$L$ is absolutely sterile, it would not decay through a virtual black hole mechanism because there 
is no way to $L$ to be transformed to spin zero particle which may be captured by the BH. To eliminate this
pessimistic possibility we can introduce an arbitrary weak coupling of $L$  to a hypothetical vector or scalar
boson $V$ with extremely weak coupling $\alpha_L$ to spin 1/2 particles $f$. 
The diagram describing the decay of $L$ through
the virtual black holes would be the same { as those} presented in Fig.~\ref{mu-3e} and \ref{mu-e-gamma}
with the substitution of $L$ instead of $\mu$, $V$ instead of $\gamma$ and $f$ instead of $e$. The decay
width of $L$ in this case case would be essentially the same as given by Eq.~(\ref{gamma-mu-3e})
with $M_L$ instead of $m_\mu$ and $\alpha_L$ instead of $\alpha$:
\be
\Gamma(L\rar 3f)_d = \frac{\alpha_L^2 M_L}{2^{11}\pi^5}\,
\left(\ln \frac{M_*^2}{M_L^2}\right)^2\,
\left(\frac{M_L}{M_*}\right)^{4\left(1+\frac{1}{d+1}\right)}
\kappa^{\frac{2}{d+1}}, 
\label{gamma-L-3e}
\ee
where the decay products, $f$, are light leptons of the Standard Model, in particular, neutrinos, and the mass $M_L$ is to be taken about $10^{12}$ GeV.
Choosing the value of $\alpha_L$
we would come to new superweakly interacting  dark matter
fermion $L$ with life-time $\tau_L = 1/\Gamma(L\rar 3f)_7$}, say, $10^3 - 10 ^4$ larger than the universe age, allowing it to make an
interesting contribution to high energy cosmic rays. To this end we need  $\alpha_L =0.028$ for $\tau_L/\tau_U = 10^3$ 
and \hbox{$\alpha_{L}=0.0088$} for $\tau_L/\tau_U = 10^4$.

As is argued in the following section such heavy leptons could be efficiently produced by the oscillating curvature in $R^2$-modified gravity.

 
 {\section{Particle production by oscillating curvature
 \label{s-R2-modified}} }

We assume that  superheavy DM particles have been created by oscillating curvature scalar 
$R (t)$ in the model of the Starobinsky inflation~\cite{Starobinsky:1980te}. The action of this theory has
the form:
\be
S (R^2) = -\frac{M_{Pl}^2}{16\pi} \int d^4 x \sqrt{-g}\,\left[R- \frac{R^2}{6M_R^2}\right].
\label{action-R2}
\ee
The non-linear term in the action leads to appearance of a new dynamical  scalar degree of freedom, 
$R(t)$, named scalaron. The parameter $M_R$
is the scalaron mass. Angular fluctuations of the cosmic microwave background radiation (CMBR) imply the following value of the scalaron mass:
$M_R \approx 3\times 10^{13} $ GeV~\cite{Faulkner:2006ub}.  As it is shown  
{in the quoted paper \cite{Faulkner:2006ub},} 
the CMBR fluctuations are expressed through the Planck and the scalaron masses in the following way:
\be
\delta ^2 \sim \left(\frac{M_R}{M_{Pl}}\right)^2, 
\label{CMBR}
\ee
where $\delta $ is the amplitude of the scalar fluctuations fixed by the observations. Thus, in models where the fundamental gravitational coupling is 
determined by $M_*$, instead of $M_{Pl}$, the scalaron mass should be changed appropriately, 
$M_R^* =   3\times 10^{13}(M_*/M_{Pl})$ GeV.  
The multidimensional gravitational action is essentially taken from ref.~\cite{Antoniadis:1998ig}, eqs. (1) and (2) with an evident change of notations.

We are interested in the case when the scalaron decays create particles with energies $10^{21}$ eV, that is the energy of UHECR. Thus, the scalaron mass, $M_R^*$,  should be {at least} of the order $10^{12}$ GeV. To this end we need to choose 
\be {
M_* = M_{Pl}/30 \approx 3 \times 10^{17}\,\,\ {\rm GeV}. }
\label{M-star} 
\ee

There are two possibilities depending upon the values to the model parameters, namely the energy
scale of $R^2$ is either larger or smaller than the scale where multidimensional gravity operates.
In our paper we assumed that the scale of $R^2$ inflation is smaller than the scale at which gravity
"lives" in extra dimensions. 
{ However, in the canonical version $R^2$ inflation operates above the spatial scales where
extra dimensions are essential. In this case a simpler situation is realised and the particle production goes
along the standard lines. 
This fact does not change our conclusions. The only essential  thing is a possibility to create
stable superheavy particles with masses around or above $10^{22}$ eV by the scalaron remains intact.}

In the considered scenario elementary particles are produced in the process of the universe heating
after Starobinsky inflation \cite{Starobinsky:1980te} induced by a high value of the curvature scalar $R(t)$. The
probabilities of different channels of this process, often called scalaron decays, are analysed and
summarised in Refs.~\cite{Arbuzova:2021etq,Arbuzova:2021oqa}. The results of these works can be
applied to our case if we substitute $M_*$ instead of $M_{Pl}$ and  $M_R^*$ instead of the scalaron mass $M_R$. 
Radiative corrections to the decay widths \cite{Arbuzova:2021etq,Arbuzova:2021oqa} are calculated in Ref.~\cite{Latosh:2022hrf}.

It is shown in Ref.~\cite{Arbuzova:2021etq} that the process of the universe heating after 
$R^2$-inflations allows for superheavy stable particles, with the interaction strength typical for 
supersymmetry, to be realistic candidates for DM. 
As is known, in the canonical model the lightest SUSY 
particle would overclose the universe if its mass is larger than several hundred GeV. This restriction is
lifted in $R^2$-gravity because the long-time heating of the universe by the scalaron decay strongly 
dilute the density of the stable superheavy particles.	
 By this reason  $R^2$-inflation opens a nice possibility for strongly interacting superheavy
 particles to make cosmological DM.    
 The width of the scalaron decay into a pair of fermions with mass $m_f$ is equal to     
  \be
  \Gamma_f = \frac{m_f^2 M_R}{6 M_{Pl}^2},
  \label{Gamma-f}
  \ee
 see e.g. Ref.~\cite{Arbuzova:2021etq,Arbuzova:2021oqa}. This result is obtained for fermions with masses much smaller than
 the scalaron mass. As it is shown in papers \cite{Arbuzova:2018apk,Arbuzova:2021etq} in the case of the dominant decay mode of the scalaron into a pair of 
 fermions with mass $m_f \sim 10^7$~GeV, the dark matter  particles, produced in secondary reactions in plasma, would have the necessary cosmological density if their mass is about $10^6$~GeV.

In this work we are interested in extremely heavy DM particles with masses comparable with the
 scalaron mass. Equation (\ref{Gamma-f}) for  the width of the scalaron decays into such superheavy 
 fermions,  $L$, is modified by the phase space factor   $\sqrt{1- 4M_L^2/M_R^2}$:
\be
 \Gamma_L = \frac{M_L^2 M_R}{6 M_{Pl}^2} \sqrt{1- \frac{4M_L^2}{M_R^2}}.
 \label{Gamma-L}
 \ee

  The superheavy fermions $L$ with mass  $M_L \sim M_R/2$ produced by the scalaron decay could form cosmological dark matter, since 
    the phase space factor diminishes their density 
    down to the observed density of DM. 
 
For the case considered in the presented work the following substitutions are to be made: $M_{Pl}\rar M_*$ and $M_R \rar M_R^*$. The decay of these heavy fermions with the mass of the order $10^{12}$ GeV through the virtual black hole could create, in particular, the usual neutrinos of very high energies. It can explain the IceCube and Baikal observations, see e.g. Refs. \cite{Halzen:2022pez} and \cite{Stasielak:2021gcv}.
\\[2mm]

\section{Conclusion \label{s-concl}}

It is argued that in the model of high dimensional gravity modification there may exist superheavy
dark matter particles stable with respect to the conventional particle interactions. However, such DM
particles should decay though the virtual black hole formation. 

We have studied two possible types of such DM particles, proton-like one consisting from three superheavy
quarks and the lepton type particle similar to superheavy neutrino. The latter option looks more
natural and less parameter dependent. An attractive feature of this version is creation of very energetic
fluxes of cosmic neutrinos, the origin of which is not well understood.    

With a proper choice of the parameters 
the life-time of such quasi-stable particles may be larger than the universe age by 3-4 orders of magnitude.
This permits them to make an essential contribution to the flux of high energy cosmic rays and, in 
particular, beyond the energies of GZK cutoff.
The considered mechanism  may lead to efficient creation of cosmic ray neutrinos of very high energies 
observed at IceCube and Baikal detectors.

\section*{Acknowledgement}
The work by E.V. Arbuzova and A.A. Nikitenko was supported by RSF grant 22-22-00294. 

\end{document}